\documentstyle[12pt,epsfig]{article}

\def\be{\begin{equation}}
\def\ee{\end{equation}}

\def\pmb#1{\setbox0=\hbox{#1}
 \kern-.025em\copy0\kern-\wd0
 \kern.05em\copy0\kern-\wd0
 \kern-.025em\raise.0433em\box0 }

\def\3{\ss}
\def\sq{\hbox{\rlap{$\sqcap$}$\sqcup$}}
\def\qed{\ifmmode\sq\else{\unskip\nobreak\hfil
\penalty50\hskip1em\null\nobreak\hfil\sq
\parfillskip=0pt\finalhyphendemerits=0\endgraf}\fi}

\def\bbbz {{\sf Z\!\!Z}}

\def\ss{\bf S}

\def\jj{\bf J}

\def\D9{\widetilde{\mbox{D}9}}
\def\D8{\widetilde{\mbox{D}8}}
\def\D7{\widetilde{\mbox{D}7}}
\def\D6{\widetilde{\mbox{D}6}}
\def\D5{\widetilde{\mbox{D}5}}
\def\D4{\widetilde{\mbox{D}4}}
\def\D3{\widetilde{\mbox{D}3}}
\def\D2{\widetilde{\mbox{D}2}}
\def\D1{\widetilde{\mbox{D}1}}
\def\D0{\widetilde{\mbox{D}0}}

\def\drawbox#1#2{\hrule height#2pt 
        \hbox{\vrule width#2pt height#1pt \kern#1pt 
              \vrule width#2pt}
              \hrule height#2pt}

\def\Fund#1#2{\vcenter{\vbox{\drawbox{#1}{#2}}}}
\def\Asym#1#2{\vcenter{\vbox{\drawbox{#1}{#2}
              \kern-#2pt       
              \drawbox{#1}{#2}}}}

\def\funda{\Fund{6.5}{0.4}}
\def\asymm{\Asym{6.5}{0.4}}

\def\symm{\funda\kern-0.4pt\funda}

\begin{document}

\begin{titlepage}

\begin{center}
\hfill CALT-68-2297 \\
\hfill CITUSC/00-052 \\
\hfill hep-th/0009252

\vskip 0.5 cm
{\Large \bf Tachyon Condensation in Unstable Type I\\
D-brane Systems}
\vskip 1 cm 
Oren Bergman\footnote{Electronic mail: bergman@theory.caltech.edu}\\   
\vskip 0.5cm

{\sl California Institute of Technology,
Pasadena, CA 91125, USA\\
and\\
CIT/USC Center for Theoretical Physics, \\
Univ. of Southern California,
Los Angeles CA}

\end{center}

\vskip 0.5 cm
\small
\begin{abstract}

Type I string theory provides eight classes of unstable D-brane
systems. We determine the gauge group and tachyon spectrum
for each one, and thereby describe the gauge symmetry breaking
pattern in the low-energy world-volume field theory.
The topologies of the resulting coset
vacuum manifolds are related to the real K-theory groups $KO^{-n}$, 
extending the known relations between the Type II classifying
spaces $BU$ and $U$ and the complex K-theory groups $K^0$ and $K^{-1}$.
We also comment on the role of the background D9-branes.

\end{abstract}

\end{titlepage}

\normalsize

\setcounter{footnote}{0}

\section{Introduction}

An unstable system of D-branes is characterized by the existence 
of tachyonic modes in the open string spectrum.
Examples of such systems include brane-antibrane configurations
in Type II string theory, and unstable D-branes in Type II and bosonic
string theory.
Since tachyons have a negative mass-squared they are
expected to condense to a non-vanishing expectation value
corresponding to the true vacuum of the open string theory, 
provided the complete potential is bounded from below.\footnote{In the 
bosonic case the potential is unbounded from below, and the condensate
is at most meta-stable.} 
On the other hand, since D-branes can also be thought of
as states in a closed string theory,
one expects an unstable system of such states to decay
into a stable closed string state carrying the same charges. 
In particular, a system carrying no net charge should decay into
the closed string vacuum.

Sen has conjectured that this decay is described by tachyon 
condensation \cite{Sen_tachyon}. 
This conjecture makes three important predictions.
First, the tachyon vacuum should be identified with the closed
string vacuum, and therefore its negative energy density
should precisely cancel the positive energy density of the unstable
D-brane system. Second, soliton configurations of the tachyon field
should be identified with lower-dimensional D-branes, and 
finally, all open string states should be removed from the spectrum in the
tachyon vacuum. 
A complete proof of the conjecture presumably requires solving the full 
open string field theory associated with the unstable D-brane system,
and there are recent indications that this may indeed be possible 
\cite{Kostelecky},\cite{BSFT}.
Notwithstanding, a considerable amount of evidence for the correctness 
of the conjecture 
has accumulated from various sources,
including low-energy world-volume field theory
\cite{Sen_action,Piljin,confine,noncomm}, 
conformal field theory \cite{CFT},
and level-truncated open string field theory \cite{SFT}.
The latter in particular has been extremely successful in testing
the first prediction, and all three have been useful in verifying
the second. The last prediction of Sen's conjecture, and probably the
most significant, is less well established. 
Whereas the first two
are essentially classical results, understanding how the open string
states are removed from the spectrum appears to require taking quantum 
effects into account \cite{confine}.

At the most basic level, an unstable D-brane system is described
by a gauge field theory, where the tachyon plays the role of an
ordinary Higgs field. The process of tachyon condensation then
reduces to the Higgs mechanism.
This simple description, while clearly not 
complete, already incorporates some of the features described above.
In particular, the Higgs mechanism creates a mass gap for some
of the fields, which can be seen as a low-energy manifestation of 
the removal of (some) open string states from the spectrum.
In addition, spontaneously broken gauge theories exhibit stable
soliton configurations, 
whose topological charges are classified by the homotopy groups of 
the vacuum manifold. These can in turn be identified with
the topological charges of lower-dimensional D-branes via K-theory 
\cite{Witten_K,Horava_K}.

The systems studied most extensively are brane-antibrane 
configurations in Type II string theory, and unstable D-branes in Type II
and bosonic string theory. The gauge group for 
$N$ brane-antibrane pairs is $U(N)\times U(N)$, and the tachyon 
transforms in the bi-fundamental representation. 
Tachyon condensation therefore breaks the group to the diagonal $U(N)$,
and the vacuum manifold is equivalent to $U(N)$.
The corresponding spectrum of stable solitons consists (for large enough
$N$) of all even co-dimension configurations, in agreement with the
spectrum of stable (BPS) D-branes.
The gauge group on a system of $N$ unstable D-branes is $U(N)$,
and the tachyon transforms in the adjoint representation.
In this case 
the unbroken gauge group (for even $N$)
is given by $U(N/2)\times U(N/2)$ \cite{Li,Horava_K}. 
The resulting vacuum manifold
is $U(N)/(U(N/2)\times U(N/2))$, which for large $N$ is equivalent to
the universal classifying space $BU$. This time the spectrum of stable 
solitons consists of odd co-dimension configurations, which again
agrees with the stable D-brane spectrum.\footnote{This is true for 
the Type II unstable D-branes, but not for the bosonic D-branes. 
Due to the cubic term in the tachyon potential for the latter, 
the gauge group $U(N)$ is actually unbroken in the bosonic case
(see \cite{Ruegg} for a discussion of the closely related $SU(N)$ case).
The vacuum manifold is therefore trivial, which is consistent with
the fact that all bosonic D-branes are unstable.}

The aim of this paper is to extend the above picture to
unstable D-brane systems in Type I string theory. 
Here we find eight classes of systems,
with different gauge groups and different tachyon spectra.
In section~2 we describe how to obtain the gauge group and 
tachyon representation in each case. In section~3 we make contact 
with real K-theory, generalizing the known relations between
complex K-theory and unstable D-brane systems in Type II.
In section~4 we discuss a couple of subtleties associated with
the background D9-branes of Type I, and we end with some conclusions
and open questions in section~5.
For convenience, the main results of the paper are summarized 
in table~1 below.
\begin{table}[htb]
\begin{tabular}{|l|l|c|l|l|}
\hline
 Type I system & ${\cal G}$ & tachyon & ${\cal H}$ & ${\cal G}/{\cal H}$ \\
\hline
\hline
 $N(\mbox{D9}+\overline{\mbox{D9}})$ & $O(N)\times O(N)$ 
  & $(\funda ,\funda )$ & $O(N)$ & $O$ \\
 $N$ D8 & $O(N)$ & $\asymm $ & $U(N/2)$ & $O/U$ \\
 $N$ D7 & $U(N)$ & $\asymm $ & $Sp(N)$ & $U/Sp$ \\
 $N$ D6 & $Sp(N)$ & $\asymm $ & $Sp(N/2)\times Sp(N/2)$ 
  & $BSp$ \\
 $N(\mbox{D5}+\overline{\mbox{D5}})$ & $Sp(N)\times Sp(N)$
  & $(\funda ,\funda )$ & $Sp(N)$ & $Sp$  \\
 $N$ D4 & $Sp(N)$ & $\symm $ & $U(N/2)$ & $Sp/U$ \\
 $N$ D3 & $U(N)$ & $\symm $ & $O(N)$ & $U/O$ \\
 $N$ D2 & $O(N)$ & $\symm $ & $O(N/2)\times O(N/2)$ & $BO$ \\
 $N(\mbox{D1}+\overline{\mbox{D1}})$ & $O(N)\times O(N)$ 
  & $(\funda ,\funda )$ & $O(N)$ & $O$ \\
 $N$ D0 & $O(N)$ & $\asymm $ & $U(N/2)$ & $O/U$\\
 $N$ D(-1) & $U(N)$ & $\asymm $ & $Sp(N)$ & $U/Sp$ \\
\hline
\end{tabular}
\caption{\small
Tachyon condensation and classifying spaces in Type I unstable D-brane
systems. $N$ is assumed to be even for the $p=8,7,5,4,2,0$ and $-1$
systems, and a multiple of 4 for $p=6$.} 
\label{t1}
\end{table}

\section{Unstable Type I D-brane systems}

In this section we shall determine the gauge group and tachyon
representation for each of the unstable D-brane systems in Type I
string theory. Some of this information, such as the gauge group
for $p=1,5$ and 9, and the tachyon spectrum for
$p=3$ and 7, can be obtained using the consistency
conditions of \cite{Gimon_Polchinski}. The rest, such as 
the gauge group and tachyon spectrum for even $p$, will
be obtained by other means.
We will ignore the background D9-branes for now, and consider their effect 
in section 4.

\subsection{$p$ odd}

The odd-dimensional systems correspond to 
D$p$-$\overline{\mbox{D}p}$ systems in Type IIB string theory
projected by $\Omega$.  
The Type IIB system has a gauge group $U(N)\times U(N)$
from the diagonal $p-p$ and $\overline{p}-\overline{p}$
sectors,
and a tachyon transforming in the bi-fundamental representation 
$({\bf N},\overline{\bf N})\oplus (\overline{\bf N},{\bf N})$
from the off-diagonal $p-\overline{p}$ and $\overline{p}-p$ sectors.

For $p=1,5$ and $9$ the RR component of the boundary state
is invariant under $\Omega$ \cite{Bergman_Gaberdiel}, so both the
D$p$-brane and the $\overline{\mbox{D}p}$-brane are invariant.
Since $p-p$ (and $\overline{p}-\overline{p}$) strings are 
therefore mapped to themselves, one can use the consistency
conditions of \cite{Gimon_Polchinski} to determine the gauge group.
One simply gets twice the result of \cite{Gimon_Polchinski},
{\em i.e.} ${\cal G}=O(N)\times O(N)$ for $p=1$ and 9, and 
$Sp(N)\times Sp(N)$ for $p=5$.
On the other hand, since $p-\overline{p}$ strings are mapped to
$\overline{p}-p$ strings, these conditions do not restrict the
tachyon spectrum. The projection simply picks out one linear 
combination of the two sectors with an arbitrary phase. 
This leaves precisely the {\em bi-fundamental}
representation regardless of the phase.
The unbroken 
group ${\cal H}$ is therefore the diagonal combination
of the two factors in ${\cal G}$.

For $p=3$ and $7$ the RR component of the 
boundary state is odd, so the brane and antibrane are interchanged. 
In this case $p-\overline{p}$ strings are mapped to themselves,
so the Gimon-Polchinski conditions fix the action of $\Omega$
on the tachyons. One finds
\be
 T_{a\overline{b}} \longrightarrow \omega_p T_{b\overline{a}}\;,
\label{tachyons}
\ee
where $\omega_p = (-i)^{1+(9-p)/2}$ \cite{Frau,Witten_K}.
The gauge group is not fixed by these conditions, but, as
with the tachyons in the previous case, the projection simply
keeps one linear combination of the two $U(N)$'s with an
arbitrary phase \cite{Witten_K}. It then follows from (\ref{tachyons})
that the tachyons transform in the {\em symmetric} representation
of $U(N)$ for $p=3$, and in the {\em antisymmetric} representation
for $p=7$.
The unbroken gauge groups in these cases were determined in \cite{Li}.
One gets $O(N)$ in the first case, and $Sp(N)$ in the second case.

\subsection{$p$ even}

For the even-dimensional branes we start with the corresponding
non-BPS D$p$-brane in Type IIB string theory, which has
a gauge group $U(N)$ and a tachyon in the adjoint representation.
The world-sheet arguments of \cite{Gimon_Polchinski} fail
in this case, since the D$p$-D9 strings 
have an odd number of mixed (ND) boundary conditions, and
therefore an odd number of fermionic zero modes in both the R
and NS sectors.
Instead we will determine 
${\cal G}$ by demanding that the D$p$-D9
strings are real. Since these strings transform in the spinor representation
of the Clifford algebra formed by the fermionic zero modes, 
and at the same time in the fundamental representation of the gauge group,
demanding reality will restrict the form of ${\cal G}$.

In the R sector there are $p+1$ fermionic zero modes (one of which
is time-like), so the ground state transforms as a spinor of 
$SO(p,1)$. In the NS sector there are $9-p$ zero modes, and the ground
state is a spinor of $SO(9-p)$. For odd $p$ there are two irreducible
representations in both sectors corresponding to the two Weyl spinors.
The zero-mode part of the GSO projection $(-1)^f=\prod\Gamma_i$ removes 
one and keeps the other.
For even $p$ there is only one irreducible spinor and no GSO 
projection.\footnote{The odd-dimensional Clifford algebras associated 
with the fermionic zero modes for even $p$ actually have {\em two}
irreducible representations (related by $\Gamma_i\rightarrow -\Gamma_i$), 
which are equivalent as representations of the corresponding  
groups. This subtlety can be overlooked in determining the
spectrum, but becomes important in the path integral, where one
is required to sum over spin structures \cite{Witten_K}.
To facilitate this sum, Witten proposed to include an additional
world-sheet fermion $\eta(\tau)$ at each boundary of the open string
ending on a D$p$-brane \cite{Witten_K}.
The GSO operator $(-1)^f=\eta\prod\Gamma_i$ then provides the
required prescription.
The signature of $\eta$ must be chosen such that the Weyl spinor
one gets after GSO-projection has the same properties as the original 
spinor. The R sector Clifford algebra is therefore increased to that
of $SO(p+1,1)$ for $p=0$ (mod 4), and $SO(p,2)$ for $p=2$ (mod 4).
In the NS sector one effectively gets $SO(9-p,1)$ in the first case,
and $SO(10-p)$ in the second.}
The resulting spinors are {\em real} for $p=0,1,2$ (mod 8), 
{\em pseudoreal} for $p=4,5,6$ (mod 8), and {\em complex} for $p=3,7$ 
(mod 8). Consequently ${\cal G}$ is {\em orthogonal} in the first case, 
{\em symplectic} in the second case, and {\em unitary} in the third case.
This is consistent with what we found in the previous section for odd
$p$. Note however that the above argument does not determine how many 
factors of these groups 
${\cal G}$ contains. In particular, for $p=1,5$ and 9 there are two factors,
whereas for $p=3$ and 7 there is only one. This is of course a consequence
of the fact that in the former case the Type I system consists of BPS branes
and BPS antibranes, and in the latter only of non-BPS branes.
By analogy, we conclude that the gauge groups for even $p$ must be
$O(N)$ for $p=0,2$ (mod 8), and $Sp(N)$ for $p=4,6$.

In order to determine the representation of the tachyon we need
to know how $\Omega$ acts on the tachyon wavefunction, and how it acts
on the CP factors of the $p-p$
string. The latter can be determined from the above results, 
together with the fact that the vector is odd under 
$\Omega$ \cite{Polchinski_II}. We find
\be
 \Lambda \longrightarrow \gamma_\Omega \Lambda^T \gamma_\Omega^{-1}\;,
 \quad
 \gamma_\Omega = \left\{
 \begin{array}{cl}
  {\bf 1} & p=0,2\;\; \mbox{(mod 8)} \\
  \jj & p=4,6 \;,
 \end{array}\right.
\ee
where
\be
 \jj = \left(
 \begin{array}{cc}
  0 & i{\bf 1} \\
  -i{\bf 1} & 0
 \end{array}
 \right)\;.
\ee
The action on the tachyon wavefunction\footnote{This is not fixed by
action on the vector since the two states belong to different
CP sectors \cite{Sen_review}.} can be determined using Sen's approach, by
considering a disc amplitude
in Type IIB string theory involving a tachyon vertex operator at the 
boundary and a RR $p$-form vertex operator in the bulk 
\cite{Sen_review}. 
Both vertex operators contribute a $\sigma_1$ factor in the trace, so the
amplitude is non-zero, and gives rise to the interaction
\be
 \int C^{(p)}\wedge dT \;.
\ee
Since $C^{(p)}$ is odd under $\Omega$ for $p=0$ (mod 4), and
even for $p=2$ (mod 4), it follows that $T$ is odd for $p=0$ (mod 4)
and even for $p=2$ (mod 4). Combining this with the action on the
CP factors gives {\em antisymmetric} representations for $p=0,6$ (mod 8),
and {\em symmetric} representations for $p=2,4$.

The unbroken gauge groups for $O(N)$ with a Higgs field
in the symmetric and antisymmetric representations were
determined in \cite{Li}, and are given by $O(N/2)\times O(N/2)$
and $U(N/2)$, respectively. The unbroken gauge groups for $Sp(N)$
can be determined in a similar way, and turn out to be
$U(N/2)$ for a symmetric Higgs, and $Sp(N/2)\times Sp(N/2)$
for an antisymmetric Higgs.

\section{Relation to real K-theory}

The topology of the tachyon vacuum manifold, {\em i.e.}
the classifying space, determines the soliton spectrum,
and therefore the spectrum of lower-dimensional D-branes.
On the other hand, since D-branes carry gauge fields they are
classified in K-theory. In particular, D-branes in Type II brane-antibrane
systems are elements of $\widetilde{K}^0(X)$, and D-branes in Type II
unstable
D-branes are elements of $K^{-1}(X)$.
It is therefore not surprising that there
exists a relation between the the mapping class groups,
{\em i.e.} homotopy groups, of the
vacuum manifold and the isomorphism class groups,
{\em i.e.} K-theory groups, of the gauge bundles.
In the complex case the relevant relations are 
\cite{Atiyah,Bott_Tu}
\begin{eqnarray}
 \widetilde{K}^0(X) & \approx & [X,BU] \nonumber \\
 K^{-1}(X) & \approx & [X,U] \;,
\label{complexK}
\end{eqnarray}
where $BU$ is the $N\rightarrow \infty$ limit of 
$U(N)/(U(N/2)\times U(N/2))$, and $U$ is the $N\rightarrow \infty$ limit of 
$U(N)$.\footnote{More generally, 
$\widetilde{K}^{-n}(X)=[X,\Omega^n BU]$, where $\Omega^n Y$
is the $n$-th iterated loop space of $Y$.  
One can prove that $\Omega BU$ is
homotopically equivalent to $U$, and $\Omega^2 BU$ is 
homotopically equivalent to $BU$.  In conjunction with (\ref{complexK}), 
this fact leads to Bott periodicity, $K^{-n-2}(X)=K^{-n}(X)$.} 
One usually assumes that the space $X$ is compact,
but an extension to non-compact $X$ is possible by defining
K-theory with compact support. 

Note that the classifying spaces
on the right hand side appear to be in the wrong place;
the vacuum manifold is $U$ for the brane-antibrane 
system and $BU$ for the unstable D-brane. 
On the other hand, non-trivial tachyon configurations 
cannot be in the vacuum everywhere in $X$, and are therefore
not classified by maps from $X$ to the vacuum manifold.
Consider for example $X={\bf R}^k$ (or its compactification ${\bf S}^k$).
Finite energy tachyon field configurations must approach the
vacuum at infinity, and are therefore associated with maps from 
${\bf S}^{k-1}$ to the vacuum manifold. These are in turn classified by
$\pi_{k-1}(U)$
for the brane-antibrane system, and by $\pi_{k-1}(BU)$ for the
unstable D-branes. From the relations\footnote{These follow from 
$\pi_k(X)=\pi_{k-1}(\Omega X)$
\cite{Bott_Tu}, and
the homotopic equivalences $\Omega BU\sim U$, $\Omega^2 BU\sim BU$.}
\begin{eqnarray}
 \pi_{k-1}(U)&=&\pi_k(BU) \nonumber \\
 \pi_{k-1}(BU)&=&\pi_k(U) \;,
\label{complexid}
\end{eqnarray}
it then follows that the tachyon vacuum manifolds appearing in 
(\ref{complexK}) should indeed be shifted by one 
relative to the corresponding K-theory.

In the real case the story is similar. The analogous relations between
the higher KO-groups and the classifying spaces are given by \cite{Karoubi}
\be
 KO^{-n}(X) \approx [X,{\cal M}_n]\;,
\ee
where, in order,
\be
 {\cal M}_n = BO,\,O,\,O/U,\,U/Sp,\,BSp,\,Sp,\,Sp/U,\,U/O 
\ee
for $n=0,\ldots,7$ (mod 8).
Since Type I D-branes in the D9-$\overline{\mbox{D9}}$ 
system (ignoring the background D9-branes) are classified
by $\widetilde{KO}^0(X)$, it is apparent from table~1 that 
the vacuum manifolds are again shifted by one
relative to the corresponding KO-groups. 
In analogy with the complex case, for $X={\bf S}^k$ this follows from 
the identities
\footnote{These follow from 
the homotopic equivalence
of the $n$-th iterated loop space of ${\cal M}_0=BO$ and ${\cal M}_n$,
{\em e.g.} $\Omega BO\sim O$, $\Omega^2 BO\sim O/U$, etc.}
\be
 \pi_{k-1}({\cal M}_{n+1}) = \pi_k({\cal M}_{n}) \;.
\ee
For example, the D5-brane can be obtained as a stable defect in the
D7-brane system, corresponding to the generator of
\be
 \pi_{1}(U/Sp) = \pi_2(O/U) = KO^{-2}({\bf S}^2) = \bbbz \;.
\ee

\section{Including the background D9-branes}

Two important subtleties arise when one includes the 32 background
D9-branes. First, since the zero-point energy of the NS sector of a 
D$p$-D9 string is given by \cite{Polchinski_II}
\be 
 m^2 = {5-p\over 8}\;,
\ee
the $p=8,7$ and 6 systems have 32 additional tachyons, which
transform in the fundamental representation of the corresponding
gauge group ${\cal G}$. From the world-volume gauge theory perspective,
their only effect is to reduce the rank of ${\cal G}$,
keeping its form fixed. The results in table~1 are therefore
qualitatively unchanged.
On the other hand, as was first pointed out in \cite{Frau},
their presence has a dramatic effect on the
single D7 and D8-brane in spacetime, which are found to be unstable
despite being tachyon-free in the $p-p$ sector. 
A natural question is what these branes decay into, given that they
carry non-trivial topological ($\bbbz_2$) charges in K-theory.

The discussion of \cite{Witten_K} seems to suggest that the
decay products are $O(32)$ gauge field configurations on the 
background D9-branes,
which are classified respectively by $\pi_1(O(32))$ and $\pi_0(O(32))$
(both of which are $\bbbz_2$).
In particular, the latter corresponds to a non-trivial 
$\bbbz_2\subset O(32)$ holonomy.
The D8-brane therefore decays into a gauge field
configuration that is pure gauge at $x=\pm\infty$, such that
$A(\infty)$ and $A(-\infty)$ are related by the reflection element 
in $O(32)$ \cite{Bergman_Schwarz}.\footnote{The Type I gauge group is 
actually $spin(32)/\bbbz_2$,
which does not contain this element \cite{Witten_K}. However,
there are two choices for the $\bbbz_2$ projection, corresponding
to which spinor chirality is kept. These are exchanged by
the reflection element of $O(32)$, so the D8-brane corresponds
to a domain wall separating regions of Type I vacuum which differ
in the chirality of the spinor state.}
Despite being topologically stable however, these gauge field
configurations are not solutions of the Yang-Mills equations.
A simple scaling argument shows that (for $p>6$) they will expand 
indefinitely to minimize the action \cite{Witten_K}, 
leaving behind a dilute gauge field.

There exist also other topologically stable gauge field configurations
corresponding to non-trivial elements in $\pi_{8-p}(O(32))$, with $p=5,1,0$
and $-1$. In contrast to the previous two, these 
will tend to shrink, which suggests the existence of $p$-dimensional
stringy objects \cite{Witten_K}. Indeed, D$p$-branes with the
above values of $p$ exist and are stable, even in the presence 
of the background D9-branes. 
Given that these D-branes are stable, what is the role of the gauge
field configurations in this case?

The answer is related to the second subtlety associated with
the background D9-branes. Namely, the gauge group of the 
unstable $p=9$ system is $O(32+N)\times O(N)$ rather than 
$O(N)\times O(N)$. The tachyon is still bi-fundamental,
so the unbroken subgroup is now $O(32)\times O(N)$.
Consequently the vacuum manifold is topologically equivalent
to $O(32+N)/O(32)$, rather than $O(N)$. For $N=1$ this is simply
the 32-sphere ${\bf S}^{32}$. More generally it is known as the real
Stiefel manifold $V_{32+N,N}$. 
A common property of this space for all $N$ is that 
$\pi_k(V_{32+N,N})=0$ for $k<32$.
Therefore there are {\em no topologically stable tachyon configurations}
(of relevance in ten dimensions). 
The D-brane charges are instead
encoded in the above gauge field configurations. The configurations
which tend to shrink ($p=-1,0,1,5$) correspond to stable D-branes,
and those which tend to expand ($p=7,8$) correspond to topologically
stable, but dynamically unstable, D-branes.

In light of this one wonders why $\widetilde{KO}^0(X)$ is the appropriate 
classifying group for D-branes in Type I string theory.
The relation to homotopy theory is
\be
 \widetilde{KO}^0({\bf S}^n) = \pi_{n-1}(O)\;,
\ee
but the vacuum manifold is actually $O(32+N)/O(32)$, not $O$.
On the other hand, as argued above, the D-brane charges are
encoded in the topology of $O(32)$, rather than the
topology of the vacuum manifold.
Since $n\leq 10$, one is well within the stability regime
for the homotopy groups of $O(32)$, which are therefore the
same as those of $O$. This is why 
$\widetilde{KO}^0(X)$ gives the correct answers.

\section{Conclusions}

Tachyons appear in many forms when supersymmetry is broken in string
theory. Since it is difficult to make general statements about the
consequence of their condensation, it is useful to have many examples.
Unstable D-brane systems provide a relatively simple arena for studying
this phenomenon for the case of open string tachyons.
In this paper we analyzed the 
unstable D-branes systems of Type I string theory in the most 
basic approach, namely the world-volume gauge theory coupled 
to a Higgs field. It would be interesting to apply some of the 
other approaches mentioned in the introduction to these systems.

Another interesting question is how the physics of unstable D-brane
systems in string theory is lifted to M-theory. 
The two brane-antibrane systems are M2-$\overline{\mbox{M2}}$
and M5-$\overline{\mbox{M5}}$.
The former is difficult to study, since
it corresponds to a non-trivial conformal field theory.
The situation is somewhat analogous to a string-antistring system,
since the M2-brane really serves as the fundamental object in
M-theory. The M5-$\overline{\mbox{M5}}$ system is, at least
qualitatively, more tractable \cite{Piljin}. 
For a single pair, it corresponds to 
a rank two antisymmetric tensor theory (without the restriction of
self-duality). 
The tachyon presumably comes from an open M2-brane, 
and therefore corresponds to a string-field. While the complete
dynamics of the M2-brane are not understood, it is straightforward
to generalize the Higgs mechanism to a rank two antisymmetric
tensor field \cite{Rey}. In particular, there exists a stable
co-dimension three defect, which can be identified with the M2-brane
\cite{Piljin}.

Another unstable system in M-theory is the non-chiral ${\bf S}^1/\bbbz_2$
compactification, also known as the $E_8\times \overline{E_8}$
system \cite{Horava_Fabinger}. In this case the two ``ends of the world''
support ordinary gauge theories, and one might expect their annihilation
to be described, in part, by an ordinary Higgs mechanism. 
In particular, since the only non-trivial configuration of $E_8$
(of relevance in ten dimensions) is the instanton associated
with $\pi_3(E_8)=\bbbz$, which is furthermore identified with the 
M5-brane, one expects the symmetry breaking pattern to be
$E_8\times E_8 \rightarrow E_8$. 
The question is whether there is a tachyon field that
achieves this.

\section*{Acknowledgment}

I would like to thank Jaume Gomis and Edward Witten
for useful conversations, and John Schwarz for earlier
collaboration on some of the material in section 4.  
I would also like to thank Edward Witten for pointing out
an error in the first version of this paper.
Finally, I thank the Aspen Center for Physics for its warm hospitality.
This work is supported in part by the DOE under grant
no. DE-FG03-92-ER 40701.


\begin{thebibliography}{[20]}


\bibitem{Sen_tachyon} 
A.~Sen,
``Tachyon condensation on the brane antibrane system,''
JHEP {\bf 9808}, 012 (1998)
[hep-th/9805170].

\bibitem{Kostelecky}
V.~A.~Kostelecky and R.~Potting,
``Analytical construction of a nonperturbative vacuum for the open  
bosonic string,''
hep-th/0008252.

\bibitem{BSFT}
A.~A.~Gerasimov and S.~L.~Shatashvili,
``On exact tachyon potential in open string field theory,''
hep-th/0009103;

D.~Kutasov, M.~Marino and G.~Moore,
``Some exact results on tachyon condensation in string field theory,''
hep-th/0009148.


\bibitem{Sen_action}
A.~Sen,
``Supersymmetric world-volume action for non-BPS D-branes,''
JHEP {\bf 9910}, 008 (1999)
[hep-th/9909062].


\bibitem{Piljin}
P.~Yi,
``Membranes from five-branes and fundamental strings from Dp branes,''
Nucl.\ Phys.\  {\bf B550}, 214 (1999)
[hep-th/9901159].


\bibitem{confine}
O.~Bergman, K.~Hori and P.~Yi,
``Confinement on the brane,''
Nucl.\ Phys.\  {\bf B580}, 289 (2000)
[hep-th/0002223];

G.~Gibbons, K.~Hori and P.~Yi,
``String fluid from unstable D-branes,''
hep-th/0009061.



\bibitem{noncomm}
K.~Dasgupta, S.~Mukhi and G.~Rajesh,
``Noncommutative tachyons,''
JHEP {\bf 0006}, 022 (2000)
[hep-th/0005006];

J.~A.~Harvey, P.~Kraus, F.~Larsen and E.~J.~Martinec,
``D-branes and strings as non-commutative solitons,''
JHEP {\bf 0007}, 042 (2000)
[hep-th/0005031];

C.~Sochichiu,
``Noncommutative tachyonic solitons: Interaction with gauge field,''
JHEP {\bf 0008}, 026 (2000)
[hep-th/0007217];

R.~Gopakumar, S.~Minwalla and A.~Strominger,
``Symmetry restoration and tachyon condensation in open string theory,''
hep-th/0007226;

G.~Mandal and S.~Rey,
``A note on D-branes of odd codimensions from noncommutative tachyons,''
hep-th/0008214;

Y.~Matsuo,
``Topological charges of noncommutative soliton,''
hep-th/0009002;

J.~A.~Harvey and G.~Moore,
``Noncommutative tachyons and K-theory,''
hep-th/0009030.


\bibitem{CFT} 
A.~Sen,
``SO(32) spinors of type I and other solitons on brane-antibrane pair,''
JHEP {\bf 9809}, 023 (1998)
[hep-th/9808141];

J.~A.~Harvey, D.~Kutasov and E.~J.~Martinec,
``On the relevance of tachyons,''
hep-th/0003101;

P.~Fendley, H.~Saleur and N.~P.~Warner,
``Exact solution of a massless scalar field with a relevant boundary interaction,''
Nucl.\ Phys.\  {\bf B430}, 577 (1994)
[hep-th/9406125];

J.~Majumder and A.~Sen,
``Vortex pair creation on brane-antibrane pair via marginal deformation,''
JHEP {\bf 0006}, 010 (2000)
[hep-th/0003124].


\bibitem{SFT} V.A.~Kostelecky and S.~Samuel,
``The static tachyon potential in the open bosonic string theory'',
Phys. Lett. {\bf B207} (1988) 169;

A.~Sen and B.~Zwiebach,
``Tachyon condensation in string field theory,''
JHEP {\bf 0003}, 002 (2000)
[hep-th/9912249];

N.~Berkovits,
``The tachyon potential in open Neveu-Schwarz string field theory,''
JHEP {\bf 0004}, 022 (2000)
[hep-th/0001084];

W.~Taylor,
``D-brane effective field theory from string field theory,''
hep-th/0001201;

N.~Moeller and W.~Taylor,
``Level truncation and the tachyon in open bosonic string field theory,''
Nucl.\ Phys.\  {\bf B583}, 105 (2000)
[hep-th/0002237];


J.~A.~Harvey and P.~Kraus,
``D-branes as unstable lumps in bosonic open string field theory,''
JHEP {\bf 0004}, 012 (2000)
[hep-th/0002117];

N.~Berkovits, A.~Sen and B.~Zwiebach,
``Tachyon condensation in superstring field theory,''
hep-th/0002211;

R.~de Mello Koch, A.~Jevicki, M.~Mihailescu and R.~Tatar,
``Lumps and p-branes in open string field theory,''
Phys.\ Lett.\  {\bf B482}, 249 (2000)
[hep-th/0003031];

P.~De Smet and J.~Raeymaekers,
``Level four approximation to the tachyon potential in superstring field  
theory,''
JHEP {\bf 0005}, 051 (2000)
[hep-th/0003220];

A.~Iqbal and A.~Naqvi,
``Tachyon condensation on a non-BPS D-brane,''
hep-th/0004015;

N.~Moeller, A.~Sen and B.~Zwiebach,
``D-branes as tachyon lumps in string field theory,''
JHEP {\bf 0008}, 039 (2000)
[hep-th/0005036];


L.~Rastelli and B.~Zwiebach,
``Tachyon potentials, star products and universality,''
hep-th/0006240;

J.~R.~David,
``Tachyon condensation in the D0/D4 system,''
hep-th/0007235;

W.~Taylor,
``Mass generation from tachyon condensation for vector fields on  
D-branes,''
JHEP {\bf 0008}, 038 (2000)
[hep-th/0008033];

R.~de Mello Koch and J.~P.~Rodrigues,
``Lumps in level truncated open string field theory,''
hep-th/0008053;

N.~Moeller,
``Codimension two lump solutions in string field theory and tachyonic  
theories,''
hep-th/0008101;

A.~Iqbal and A.~Naqvi,
``On marginal deformations in superstring field theory,''
hep-th/0008127.


\bibitem{Witten_K} 
E.~Witten,
``D-branes and K-theory,''
JHEP {\bf 9812}, 019 (1998)
[hep-th/9810188].

\bibitem{Horava_K}
P.~Horava,
``Type IIA D-branes, K-theory, and matrix theory,''
Adv.\ Theor.\ Math.\ Phys.\  {\bf 2}, 1373 (1999)
[hep-th/9812135].

\bibitem{Li}
L.~Li,
``Group Theory Of The Spontaneously Broken Gauge Symmetries,''
Phys.\ Rev.\  {\bf D9}, 1723 (1974).

\bibitem{Ruegg}
H.~Ruegg,
``Extremas Of SU(N) Higgs Potentials And Symmetry Breaking Pattern,''
Phys.\ Rev.\  {\bf D22}, 2040 (1980).


\bibitem{Gimon_Polchinski}
E.~G.~Gimon and J.~Polchinski,
``Consistency Conditions for Orientifolds and D-Manifolds,''
Phys.\ Rev.\  {\bf D54}, 1667 (1996)
[hep-th/9601038].

\bibitem{Bergman_Gaberdiel}
O.~Bergman and M.~R.~Gaberdiel,
``A non-supersymmetric open-string theory and S-duality,''
Nucl.\ Phys.\  {\bf B499}, 183 (1997)
[hep-th/9701137].

\bibitem{Frau}
M.~Frau, L.~Gallot, A.~Lerda and P.~Strigazzi,
``Stable non-BPS D-branes in type I string theory,''
Nucl.\ Phys.\  {\bf B564}, 60 (2000)
[hep-th/9903123];

M.~Frau, L.~Gallot, A.~Lerda and P.~Strigazzi,
``Stable non-BPS D-branes of type I,''
hep-th/0003022.


\bibitem{Polchinski_II}
J.~Polchinski,
``String Theory, Vol. II'',
(Cambridge, 1998).

\bibitem{Sen_review}
A.~Sen,
``Non-BPS states and branes in string theory,''
hep-th/9904207.


\bibitem{Atiyah}
M.F.~Atiyah,
``K-Theory'', (Benjamin, 1964).

\bibitem{Bott_Tu}
R.~Bott and L.W.~Tu,
``Differential Forms in Algebraic Topology'',
(Springer, 1982).

\bibitem{Karoubi}
M.~Karoubi,
``K-Theory. An Introduction'', (Springer, 1978).




\bibitem{Bergman_Schwarz} 
O.~Bergman and J.H.~Schwarz, unpublished;

J.~H.~Schwarz,
``Remarks on non-BPS D-branes,''
Class.\ Quant.\ Grav.\  {\bf 17} (2000) 1245
[hep-th/9908091].

\bibitem{Rey}
S.~Rey,
``The Higgs Mechanism For Kalb-Ramond Gauge Field,''
Phys.\ Rev.\  {\bf D40}, 3396 (1989).


\bibitem{Horava_Fabinger}

M.~Fabinger and P.~Horava,
``Casimir effect between world-branes in heterotic M-theory,''
Nucl.\ Phys.\  {\bf B580}, 243 (2000)
[hep-th/0002073].




\end{thebibliography}
\end{document}